\documentclass[3p,twocolumn,sort&compress, fleqn]{elsarticle}
\usepackage{amsmath, bm}
\usepackage{amssymb}
\usepackage{braket}
\usepackage{listings}
\usepackage{cprotect}
\usepackage{lmodern}
\usepackage{xspace}
\usepackage{ulem}
\usepackage{color}

\journal{Computer Physics Communications}

\newcommand{\ham}{\mathcal{H}}

\newcommand{\tenes}{\texttt{tenes}\xspace}
\newcommand{\tenesstd}{\texttt{tenes\_std}\xspace}
\newcommand{\tenessimple}{\texttt{tenes\_simple}\xspace}

\usepackage{hyperref}
\hypersetup{
        colorlinks=true,
}
\usepackage{listings}

\long\def\beginmypgfpdfnamed#1#2\endmypgfpdfnamed{\includegraphics{#1}}

\graphicspath{{./pyfig/}}

\begin{document}

\begin{frontmatter}

\title{TeNeS: Tensor Network Solver for Quantum Lattice Systems}

\author[issp]{Yuichi Motoyama}\ead{y-motoyama@issp.u-tokyo.ac.jp}
\author[utokyo,presto]{Tsuyoshi~Okubo}
\author[issp]{Kazuyoshi~Yoshimi}
\author[issp]{Satoshi~Morita}
\author[issp]{Takeo~Kato}
\author[issp,trans-scale]{Naoki~Kawashima}

\address[issp]{The Institute for Solid State Physics, The University of Tokyo, Chiba 277-8581, Japan}
\address[utokyo]{Institute for Physics of Intelligence, The University of Tokyo, Tokyo 113-0033, Japan }
\address[presto]{JST, PRESTO, Honcho 4-1-8, Kawaguchi, Saitama, 332-0012, Japan }
\address[trans-scale]{Trans-scale Quantum Science Institute, The University of Tokyo, Bunkyo, Tokyo, 113-0033 Japan}

\begin{abstract}
TeNeS (Tensor Network Solver)~\cite{TeNeS_webpage, TeNeS_Github} is a free/libre open-source software program package for calculating two-dimensional many-body quantum states based on the tensor network method and the corner transfer matrix renormalization group (CTMRG) method. This package calculates ground-state wavefunctions for user-defined Hamiltonians and evaluates user-defined physical quantities such as magnetization and correlation functions. For certain predefined models and lattices, there is a tool that makes it easy to generate input files. TeNeS uses an OpenMP/MPI hybrid parallelized tensor operation library and thus can perform large-scale calculations using massively parallel machines.
\end{abstract}


\end{frontmatter}
\noindent
{\bf PROGRAM SUMMARY}

  %Delete as appropriate.

\begin{small}
\noindent
{\em Program Title:} TeNeS \\
{\em Journal Reference:}                                      \\
  %Leave blank, supplied by Elsevier.
{\em Catalogue identifier:}                                   \\
  %Leave blank, supplied by Elsevier.
{\em Licensing provisions:} GNU General Public License version 3\\
{\em Programming language:} \verb*#C++# and  \verb*#python3#\\
{\em Computer:} PC, cluster machine\\ 
  %Computer(s) for which program has been designed.
{\em Operating system:} UNIX like system, tested on Linux and macOS\\ 
  %Operating system(s) for which program has been designed.
  % Please give some freely chosen keywords that we can use in a
  % cumulative keyword index.
% {\em Classification:} 4.12 Other Numerical Methods, 6 Computer Languages, Hardware and Software, 7.7 Other Condensed Matter inc. Simulation of Liquids and Solids\\ 
  %Classify using CPC Program Library Subject Index, see (
  % http://cpc.cs.qub.ac.uk/subjectIndex/SUBJECT_index.html)
  %e.g. 4.4 Feynman diagrams, 5 Computer Algebra. 7.7 Other Condensed Matter inc. Simulation of Liquids and Solids
{\em External routines/libraries:} mptensor \\
{\em Nature of problem:} 
TeNeS calculates the approximate ground states and their properties of user-defined two-dimensional quantum lattice models using user-friendly input files. Numerically exact solutions of such tasks generally require an exponentially diverging computational time, whereas the error in the output of TeNeS is well controlled and can be reduced with a polynomial cost. \\
{\em Solution method:} TeNeS implements the tensor networks method based on a tensor-product-state (TPS) wavefunction and the corner transfer matrix renormalization group method.
TeNeS also supports massively parallel computing using the library mptensor, which implements OpenMP/MPI hybrid parallelized tensor operations.\\
%{\em Running time:} $<$ 1 min\\
\end{small}

\section{Introduction}
\label{sec:intro}
In the last 20 years, several new concepts and techniques have been proposed for tensor networks (TN), and TN has become a powerful and important language in the field of quantum many-body physics~\cite{Orus2014, Orus2019}.
One of the most important applications of the TN method is ground state calculations by the variational principle. 
In TN, a quantum many-body state (wave function) is represented by a partial contraction of networks of local tensors (tensor network states, TNS).
Then, a ground state calculation is reduced to an optimization problem of finding the TNS representation with the lowest energy.
A ground state calculation using TNS has several advantages over other conventional numerical methods.
And compared with exact diagonalization, it has a much lower size restriction.
If the system has spatial translational symmetry, some implementations of the TN method allow the calculation of physical quantities in the thermodynamic limit directly by  enforcing translational symmetry in the tensor networks.
It is also notable that the TN method can treat frustrated quantum lattice models and/or fermionic systems for which the infamous sign problem appears, by employing the quantum Monte Carlo method.
The key concept in understanding the efficiency of the TN method is the area law of the entanglement entropy ~\cite{Srednicki1993,Xiang2001,Calabrese2004}, which suggests that many interesting classes of states such as ground states of gapfull lattice systems can be captured efficiently and systematically by some classes of TNS, such as matrix product states (MPS)~\cite{Fannes1992,Klumper1991,Klumper1993} and tensor product states (TPS)~\cite{Nishino2001} (also known as projected entangled pair states  (PEPS)~\cite{Verstraete2004}).

For two-dimensional lattice models, TPS and their infinite version (iTPS/iPEPS) are widely used to represent low-energy states. (See Fig.\ref{fig:iTPS} for iTPS.)
To evaluate the properties of a quantum state represented by an iTPS, we often need to contract the infinite tensor network. For this purpose, several algorithms have been developed such as the corner transfer matrix renormalization group (CTMRG) ~\cite{Baxter1968,Baxter1978,Nishino1996,Orus2009, Phien2015, Lee2018}, the boundary MPS algorithm~\cite{Jordan2008, Xie2017}, and the (cluster) mean-field approximation~\cite{Picot2015, Picot2016, Jahromi2019, Jahromi2020, Vlaar2021}.
To optimize an iTPS for a given Hamiltonian, the imaginary time evolution~\cite{Jiang2008, Li2012, Jiang2008, Jordan2008, Orus2009, Phien2015} and variational update~\cite{Corboz2016, Vanderstraeten2016,Liao2019} are typically used.
Recently, extensions to low-lying excited states~\cite{Vanderstraeten2019, Ponsioen2020, Boris2020},  finite temperature properties~\cite{Czarnik2012, Czarnik2015, Czarnik2016, Kshetrimayum2019, Wietek2019}, and real-time evolution~\cite{Czarnik2019, Kshetrimayum2020} have also been discussed.
The iTPS method has been applied to several two-dimensional quantum lattice systems such as a variety of square lattice models~\cite{Niesen2017,Haghshenas2018, Yamaguchi2018}, a triangular lattice magnet~\cite{Niesen2018}, kagome magnets~\cite{Picot2015,Liao2017, Lee2018, Okuma2019}, the Shastry--Sutherland model~\cite{Corboz2013,Corboz2014_2}, Kitaev models~\cite{Iregui2014, Okubo2017,Lee2020,  Lee2020a}, itinerant electron systems~\cite{Corboz2011, Corboz2014, Corboz2016_2, Boris2020}, and hardcore Bose--Hubbard models~\cite{Jordan2009, Tu2020,Wu2020}.

Developing a TPS-based numerical application program requires implementing basic tensor-manipulations in our code.
Fortunately, for this purpose a number of libraries for tensor operations are available in modern scientific programming language.
For example, we can use ITensor~\cite{itensor, itensor_web}, TeNPy~\cite{tenpy, tenpy_web}, Uni10~\cite{uni10, uni10_web}, SyTen~\cite{syten, syten_web}, TensorNetwork~\cite{tensornetwork, tensornetwork_web}, TiledArray~\cite{tiledarray_web}, Cyclops~\cite{Cyclops, Cyclops_web}, mptensor~\cite{mptensor, mptensor_Github}, PyTorch~\cite{pytorch, pytorch_web}, and TensorOperations.jl~\cite{tensor_operations_jl_web} (for detailed information, see, e.g., the web site~\cite{tensornetwork_org}).
In addition, many algorithm developers have published implementations that reproduce their benchmark results.
To the best of our knowledge, however, ready-to-use solver packages based on the iTPS algorithm have not been provided to date.
Many published codes apply to square lattices, and adapting other lattices, such as triangular and kagome lattices, is not straight-forward.
Although the implementation of the algorithms is not difficult thanks to the intuitive graph notations and sophisticated programming tools/libraries, it is still time-consuming to develop solver programs suited for product-runs in varying conditions.

Against this background, we have developed TeNeS (\textbf{Te}nsor \textbf{Ne}twork \textbf{S}olver)~\cite{TeNeS_webpage, TeNeS_Github, pTNS}, a free/libre and open-source software (FLOSS) package for finding the ground state of two-dimensional quantum lattice systems in the thermodynamic limit.
TeNeS can treat bosonic or spin systems.
In TeNeS, the iTPS ansatz is employed to express the many-body wave function of an infinite system and is optimized using the imaginary-time evolution method. The CTMRG method and the MFE method are used for contraction of the tensors.
For elementary tensor operations, TeNeS uses the tensor arithmetic library mptensor~\cite{mptensor_Github,mptensor}, which is developed by one of the authors of this paper.
Although TeNeS always considers iTPS on a square lattice, it can deal with other lattices by considering middle-range interactions beyond the nearest neighbors, as described below.
We believe that TeNeS will assist users in making theoretical studies of two-dimensional quantum many-body systems using the TN method.

In this paper, we introduce TeNeS version 1.2.0.
The paper is organized as follows.
We describe the algorithms implemented in TeNeS in Sec.~\ref{sec:algorithm} and present the structure of the TeNeS code in Sec.~\ref{sec:structure}.
Sections~\ref{sec:install} and \ref{sec:usage} are devoted to the installation and usage of TeNeS, respectively.
Tutorials using typical models are presented in Sec.~\ref{sec:applications}.
Finally, we summarize our work in Sec.~\ref{sec:summary}.

\section{Algorithm of TeNeS} 
\label{sec:algorithm}

We will briefly describe the algorithms used in TeNeS in this section.

\subsection{iTPS}
\label{sec:alg_iTPS}

\begin{figure}
  \centering
  \includegraphics[width=\columnwidth]{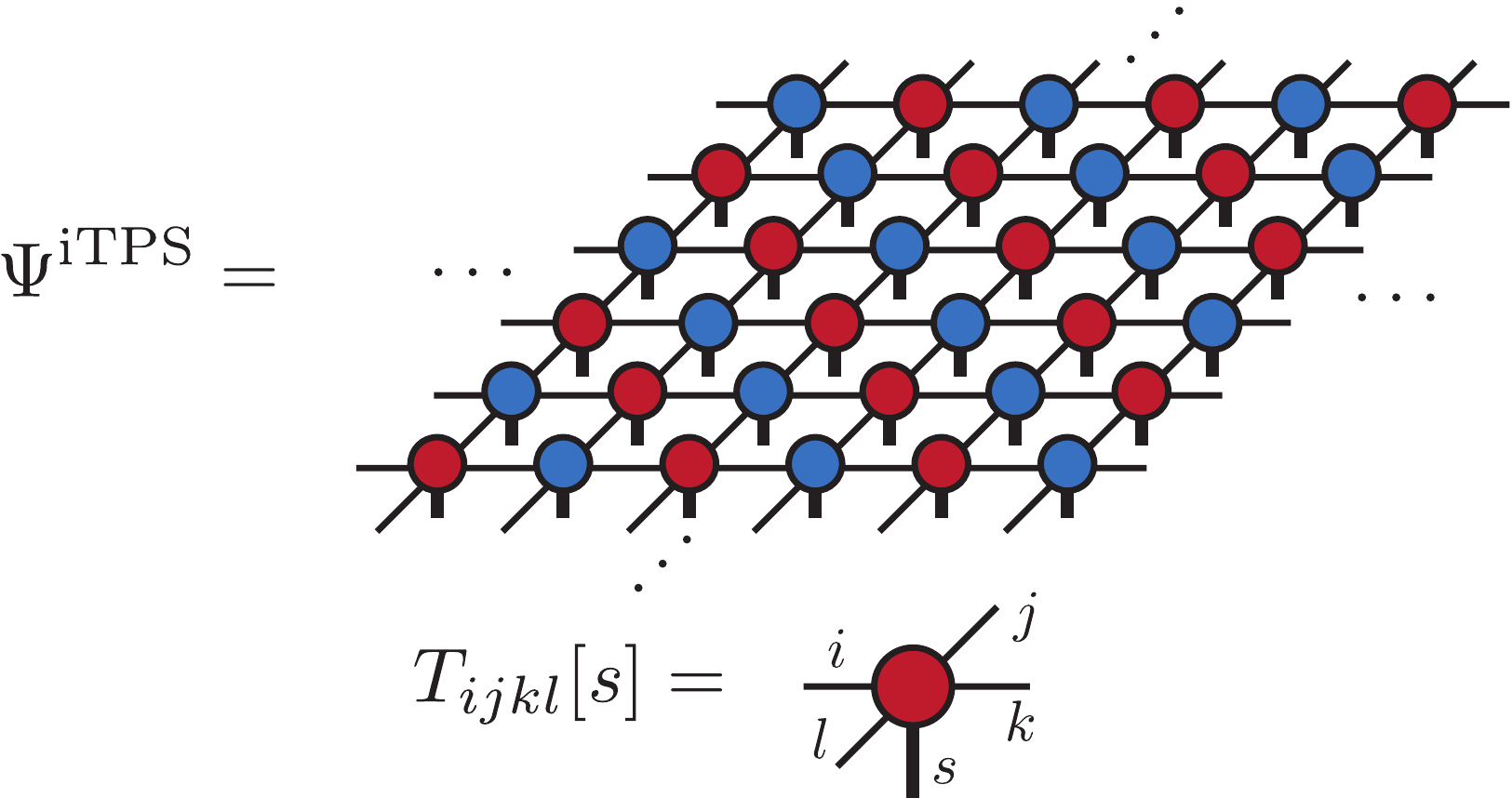}
  \caption{Tensor network diagram of iTPS in an infinitely large square lattice. Here we assume two-site periodicity in the wavefunction. Tensors with the same color have common tensor elements. 
    }
  \label{fig:iTPS}
\end{figure}

Let us consider a quantum system on a square lattice as an example.
We illustrate a tensor network diagram of iTPS for this model in Fig.~\ref{fig:iTPS}.
The circles are local tensors each having several indices in general.
The bold lines represent indices of physical bases such as states of local spins, while the thin lines indicate virtual indices connecting neighboring tensors.
The accuracy of iTPS is controlled by the bond dimension $D$, which is the dimension of the virtual indices ($i,j,k,l$ in Fig.~\ref{fig:iTPS}).
For a larger $D$, we can approximate the target wavefunction more accurately.
Typically, we are interested in the ground state of a given Hamiltonian.
It is known that, in many cases of interest, 
the entanglement entropy of the ground state wavefunction satisfies the area law~\cite{Srednicki1993, Xiang2001, Calabrese2004}, in contrast to the generic quantum states.
This fact indicates that iTPS can approximate many ground states of infinitely large systems well, even with a finite bond dimension $D$.

\subsection{Contraction of iTPS}
\label{sec:alg_contraction}
\begin{figure}
  \centering
  \includegraphics[width=\columnwidth]{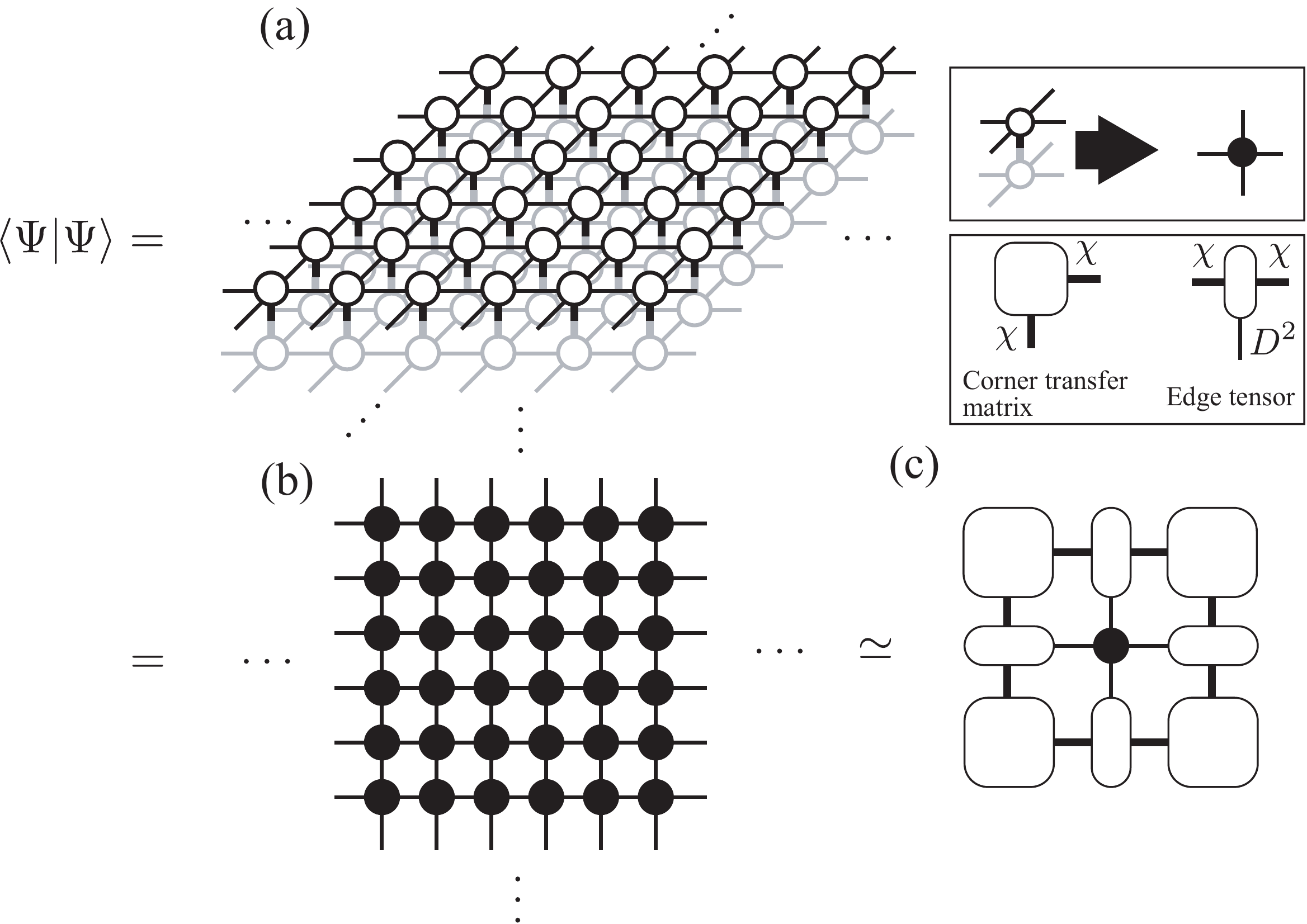}
  \caption{Tensor network diagram representing the norm of an iTPS. (a) Double-layered tensor network. (b) Single-layered classical tensor network. (c) Corner transfer matrix representation of the environment surrounding the center tensor. Note that the bond dimension of the single-layered classical tensor network is $D^2$. We represent the bond dimension of the corner matrix as $\chi$. Using CTMRG method, we iteratively determine the corner matrices and the edge tensors.}
  \label{fig:CTMRG}
\end{figure}

For a given iTPS, we can calculate the expectation values of physical quantities by 
contracting double-layered tensor networks formed by two iTPSs with the physical indices of the bra-iTPS connected to those of the ket-iTPS, as illustrated in Fig.\ref{fig:CTMRG}(a).
For this contraction, TeNeS offers two algorithms, the CTMRG method and the mean-field environment (MFE) method, either of which can be used.
In the CTMRG method~\cite{Baxter1968, Baxter1978, Nishino1996, Orus2009, Corboz2014, Phien2015},
TeNeS regards the two corresponding elementary tensors, 
one in the bra and the other in the ket, as a single tensor (Fig.\ref{fig:CTMRG}(b)) 
and constructs the corner transfer matrices and edge tensors (Fig.~\ref{fig:CTMRG}(c)) from the resulting single-layer tensor network.
The accuracy of the CTMRG method is controlled by the bond dimension $\chi$ of the corner matrices.
Since the single-layered tensor network has the bond dimension $D^2$,
$\chi$ is usually chosen to be scaled as $\chi \propto D^2$ as $D$ increases, so as to match the accuracy of the contraction to that of the ansatz.

We slightly modify the CTMRG method for iTPS with an arbitrary unit cell structure proposed in Ref.~\cite{Corboz2014}.
In the original method, QR decomposition is used to construct the projector for renormalizing the environment tensors.
However, we can avoid this step without loss of accuracy.
Furthermore, using a partial singular value decomposition (SVD) technique, such as the Arnoldi method and randomized SVD \cite{Halko2011}, we can perform SVD without explicitly making the matrix.
By combining these two modifications, we can reduce the leading computational cost of the CTMRG to $O(D^{10})$ from the original $O(D^{12})$. 

The MFE method~\cite{Picot2015, Picot2016, Jahromi2019, Jahromi2020, Vlaar2021} 
is another convenient method for contracting iTPS.
In this method, we focus on one tensor $T_{ijkl}[s]$ (ket) and its conjugate (bra), and take the effect of the other tensors into account via environment tensors (matrices) connected by virtual bonds.
These MFE tensors are calculated by the singular value decomposition in the simple update method.
Finally, these approximate tensors are contracted into one scalar with a computational cost of $O(D^4) $, which is much cheaper than that of the CTMRG method.
The contraction of larger clusters requires more computational cost, for example, $O(D^5)$ for a $1 \times 2$ cluster.
It should be noted that the MFE method does \textit{not} satisfy the variational principle, while the CTMRG method does satisfy it within the degree of the approximation controlled via the bond dimension $\chi$.
Therefore, care must be taken when comparing energies obtained by the mean-field environment.

\subsection{Optimization of iTPS}
\label{sec:alg_optim}
To optimize the tensors for the ground state of a given Hamiltonian $\mathcal{H}$, TeNeS uses imaginary time evolution (ITE):
\begin{equation}
    |\Psi^{\mathrm{iTPS}}\rangle \simeq e^{-T\mathcal{H}}|\Psi_0\rangle,
\end{equation}
where $|\Psi_0\rangle$ is an initial wave function and $T$ is a sufficiently long imaginary time.
In practice, the imaginary time evolution operator is decomposed as a product of small pieces by Suzuki--Trotter decomposition~\cite{Suzuki1976, Trotter1959}.
In TeNeS, the Hamiltonian is assumed to be represented as a sum of short-range two-body interactions as
\begin{equation}
    \mathcal{H} = \sum_{\{(i,j)\}} H_{ij}.
\end{equation}
Then, using Suzuki--Trotter decomposition, the ITE is represented as
\begin{equation}
    e^{-T\mathcal{H}} |\Psi_0\rangle= \left(\prod_{\{(i,j)\}} e^{-\tau H_{ij}}\right)^{N_\tau} |\Psi_0\rangle+ O(\tau),
\end{equation}
where $N_\tau = T/\tau$ is the number of ITEs with a sufficiently small $\tau$.
To keep the bond dimension constant along the ITE, we truncate the bond dimension to $D$, after multiplying $e^{-\tau H_{ij}}$ by the iTPS.
In TeNeS, we can use the standard simple update and full update methods for such truncations~\cite{Jiang2008, Jordan2008, Orus2009, Phien2015}.
In these methods, the bond dimension is truncated so as to minimize the norm of difference,
\begin{equation}
\| \Ket{\Psi^\text{iTPS}_\text{new}} - e^{-\tau \mathcal{H}_{ij}}\Ket{\Psi^\text{iTPS}} \|.
\end{equation}
To evaluate it, the simple update and full update method use the mean field environment method and the CTMRG method in order to contract iTPS, respectively.
TeNeS performs $N_\tau^\text{simple}$ steps simple updates and then $N_\tau^\text{full}$ steps full updates in the optimization process.

\begin{figure}
  \centering
  \includegraphics[width=\columnwidth]{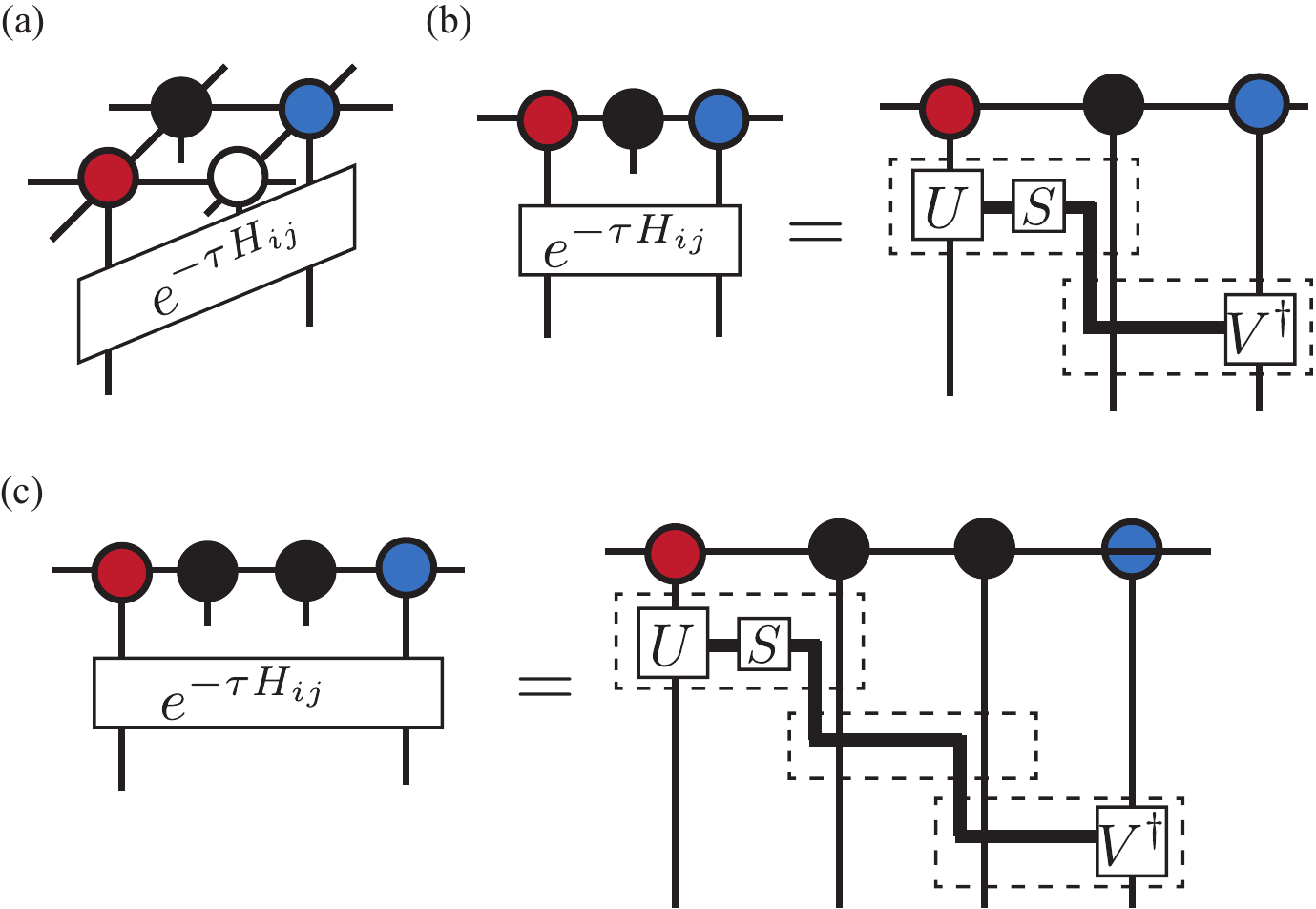}
  \caption{Example decomposition of a long-range interaction into a product of nearest-neighbor interactions in TeNeS. (a) Local structure of iTPS and an ITE operator corresponding to a ``next-nearest-neighbor'' interaction. (b) Decomposition of the ITE operator into a product of two nearest-neighbor operators. Firstly, the ITE operator is decomposed by SVD: $e^{-\tau H_{ij}}= U S V^\dagger$. To increase the accuracy of the subsequent truncation, the singular value is absorbed into the ``first layer''. Here we neglected several virtual legs for better visualization. (c) Similar decomposition in the case of two intermediate tensors.}
  \label{fig:long_range}
\end{figure}

For an interaction with a range longer than the nearest neighbors of the square lattice, 
we consider a path on the square lattice connecting the two interacting sites and 
we decompose the ``long-ranged'' ITE operator into a sequence of nearest-neighbor operators, each acting on a nearest-neighbor pair along the path (Fig.~\ref{fig:long_range}).
For the decomposition, we first SVD decompose the ITE operator.
Then, one of the unitary operators and the singular values form the ``first layer'' operator, while the other unitary operator is included in the last layer operator. (Fig.~\ref{fig:long_range} (b) and (c).)
For the intermediate layers, the nearest-neighbor operator is an ``identity'' operator
that simply passes the information as illustrated as the middle dashed square in Fig.~\ref{fig:long_range} (c).
After this decomposition to nearest-neighbor operators, we apply the standard simple update or full update methods to them.

For many specific cases, this strategy is expected to be less efficient than other techniques, such as a simple update with successive SVDs~\cite{Corboz2010, Okubo2017} or a full update based on a larger cluster~\cite{Corboz2013}.
However, we adopt this method for its flexibility, since it can be used for any long-ranged interactions, and once the decomposition of the ITE operator is given, we can in principle efficiently deal with arbitrary long-range interactions without drastically increasing the computation cost, which is not straightforward in the case of previous techniques.
In the case of two-dimensional lattices, the decomposition might become complicated,
but in TeNeS, a proper decomposition is automatically calculated.

\section{Structure of TeNeS}
\label{sec:structure}

\begin{figure}
  \centering
  \includegraphics[width=\columnwidth]{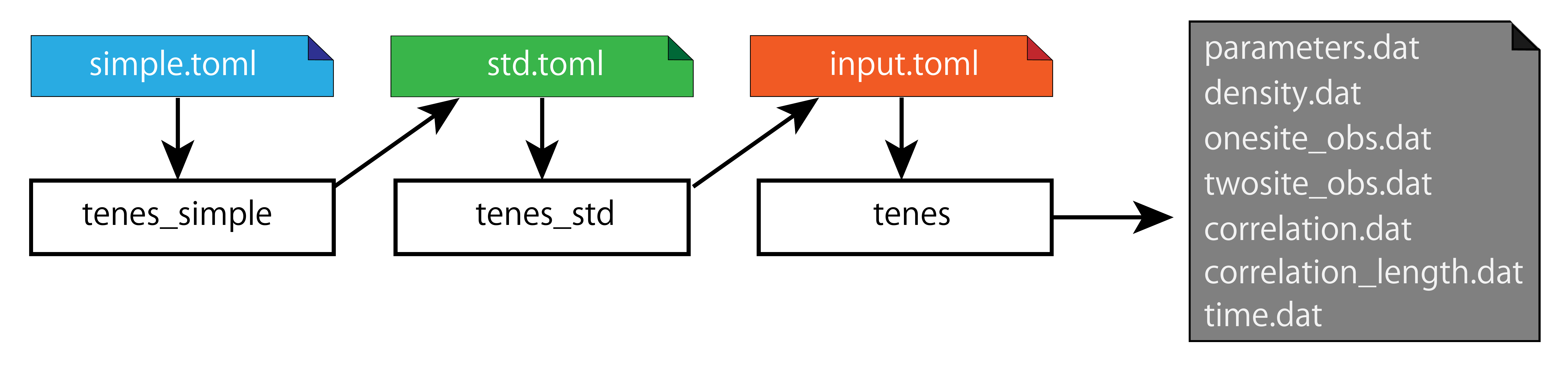}
  \caption{
    Calculation flow of TeNeS.
    The main program \tenes takes an input file and outputs calculation results to several files.
    The main program accepts an input file in the form of TOML.
    The support scripts \tenessimple and \tenesstd generate input files for \tenesstd and \tenes, respectively.
  }
  \label{fig:flow_tenes}
\end{figure}

A schematic of the calculation flow of TeNeS is shown in Fig.~\ref{fig:flow_tenes}.
The main program of TeNeS (\tenes) optimizes an iTPS based on the ITE method and evaluates the expectation value of physical quantities.
Its input file defines a model, observables, calculation parameters, etc.
By hand-writing the input files, we can in principle deal with any two-dimensional two-body interaction model.
However, for ease of use, the following scripts are provided for automatic generation of the input files: 
\begin{itemize}
    \item \tenesstd: A tool that calculates the imaginary-time evolutional operators from the Hamiltonian and generates an input file to execute \tenes.
    An input file of \tenesstd defines the lattice model etc. according to a predetermined format.
    \item \tenessimple: A tool that generates an input file for \tenesstd from an even simpler input file which specifies a predefined lattice model.
\end{itemize}

The input file of \tenessimple is similar to those of the numerical package of exact diagonalization H$\Phi$~\cite{Kawamura2017} and variational Monte Carlo methods mVMC~\cite{Misawa2019}, both developed by some of the present authors under the same software development project, PASMUS \cite{PASUMS}.
The script program \tenessimple supports simple lattice models like the XXZ model and Bose--Hubbard model on a square lattice, a triangular lattice, a honeycomb lattice, and a kagome lattice.
\tenessimple also defines some elementary observables such as the spin operator $S_i^\alpha$ and the short-range spin correlation $S_i^\alpha S_j^\alpha$.

A user can simulate other models and/or lattices than those predefined in \tenessimple by directly creating an input file of \tenesstd.
Other observable operators and the numerical condition of the two-site correlation functions than those \tenessimple prepares can be specified by editing the \tenes input.
We will describe the usage of these programs and the format of input files in more detail in Sec.~\ref{sec:usage}.

The main program \tenes outputs its calculation results into the \verb|output| directory automatically generated in the working directory.
Each file contains the energy per site, the expectation values of one-site operators, nearest-neighbor correlations between one-site operators, correlation functions along with horizontal and vertical directions, correlation length, calculation time, and calculation parameters.

\begin{figure}
  \centering
  \includegraphics[width=\columnwidth]{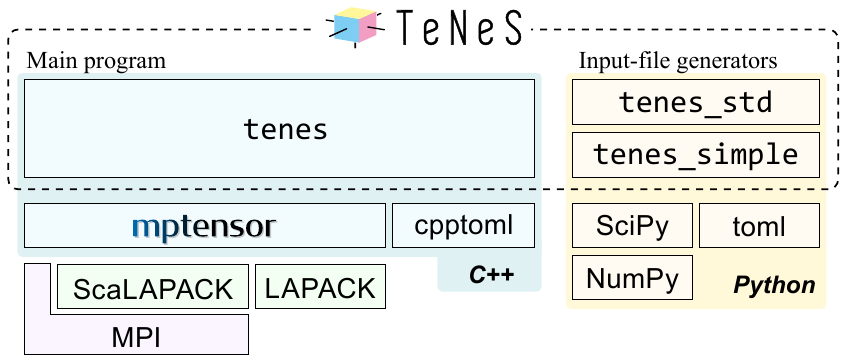}
  \caption{
    Software dependency of TeNeS. 
    Tensor operations in \tenes are based on mptensor, which uses ScaLAPACK and LAPACK.
  }
  \label{fig:tenes_structure}
\end{figure}

Figure \ref{fig:tenes_structure} shows the software dependency of TeNeS.
Tensor operations in \tenes are implemented using mptensor, which is a \verb|C++| library for parallel computations of tensor networks~\cite{mptensor}.
This library provides various tensor operations, including decomposition and contraction, commonly used in various tensor network methods.
Its interface is designed to be compatible with NumPy, a Python module for multi-dimensional arrays~\cite{numpy,numpy_web}, for ease of translation of the serial code written with Numpy to a parallelized \verb|C++| code.
By default, mptensor uses ScaLAPACK, a library for numerical linear algebra on distributed-memory machines~\cite{scalapack,scalapack_web}.
For a system without MPI, mptensor also supports LAPACK~\cite{lapack,lapack_web}.
The main object of mptensor is the \verb|Tensor| template class, which corresponds to \verb|ndarray| in NumPy.
The \verb|Tensor| class converts a tensor into a matrix, for example, $T_{ijkl}=M_{(ij)(kl)}$, and its elements are stored in a \verb|Matrix| object.
The \verb|Matrix| class is a wrapper class of linear-algebra libraries.
This class manages matrix elements distributes in a block-cyclic manner as ScaLAPACK.

\section{Installing TeNeS}
\label{sec:install}

The source code of TeNeS can be downloaded from the GitHub repository~\cite{TeNeS_Github}.
The main program of TeNeS, \tenes, is written in \verb|C++11|,
and depends on the following FLOSS packages:
\begin{itemize}
    \item mptensor~\cite{mptensor} for operating tensors in a cluster machine.
    \item cpptoml~\cite{cpptoml} for reading/parsing input files.
\end{itemize}
These are downloaded automatically through the build process.
If one wants to use MPI parallelization, MPI and ScaLAPACK are required.
Building TeNeS requires CMake version 3.6 or later.
The utility tools of TeNeS \tenessimple and \tenesstd for generating and converting input files are written in Python3.
To use these tools, a user needs to install NumPy~\cite{numpy,numpy_web}, SciPy~\cite{scipy,scipy_web}, and toml~\cite{toml_python} packages.

TeNeS can be built by the following commands:
\begin{verbatim}
$ cd TeNeS
$ mkdir build
$ cd build
$ cmake -DCMAKE_INSTALL_PREFIX=<path> ../
$ make
\end{verbatim}
\texttt{<path>} specifies the path to the directory where TeNeS will be installed (the default is  \texttt{/usr/local}).
The test suite of TeNeS runs by the \texttt{ctest} command.
Finally, the command
\begin{verbatim}
$ make install
\end{verbatim}
installs \tenes, \tenesstd, and \tenessimple into the \texttt{<path>/bin} and copies sample files into the \texttt{<path>/share/tenes/sample}.

\section{Using TeNeS}
\label{sec:usage}

We explain how to use \tenessimple, \tenesstd, and \tenes individually in the following three sections.
These programs accept an input file
in the TOML format~\cite{TOML}, where
a parameter and its value are specified as \texttt{parameter = value}.
The parameters can be categorized by a \textit{table} in the TOML format, called a ``section'' hereafter.
For example, the parameters \texttt{type} and \texttt{J} in the \texttt{model} section can be specified as follows:
\begin{verbatim}
[model]
type = "spin"
J = 1.0
\end{verbatim}
The section name surrounded by double square bracket, like \texttt{[[hamiltonian]]}, means that multiple sections with the same name can be defined simultaneously (referred to as an \textit{array of table} in the TOML format).

\subsection{Usage of \tenessimple}

\tenessimple is a tool that creates an input file for \tenesstd for predefined models and lattices.
From information in two sections, \verb|[model]| and \verb|[lattice]|, \tenessimple generates the necessary information on models and lattices in the form appropriate for the input file of \tenesstd. 
The parameters specified in the \verb|correlation| and \verb|parameter| sections are not used in \tenessimple and are just copied to the input file of \tenesstd.

In the \verb|[model]| section, the model and model parameters in the Hamiltonian are specified. For the model specification, a user chooses either a spin model or a boson model. For a spin model, the Hamiltonian that can be specified as an input to \tenessimple is given as
\begin{align}
\mathcal{H} &= \sum_{\langle ij \rangle}\left[\sum_\alpha^{x,y,z} J^\alpha_{ij} S^\alpha_i S^\alpha_j + B \left(\vec{S}_i\cdot\vec{S}_j\right)^2 \right] \\
& \hspace{5mm} - \sum_i \vec{h}\cdot \vec{S}_i - \sum_i D \left(S^z_i\right)^2,
\end{align}
where $S_i^{\alpha}$ is a spin operator with $\alpha~(= x,y,z)$ at the $i$-th site, $\vec{S}_i\equiv(S_i^x, S_i^y, S_i^z)$ is a spin operator vector at the $i$-th site, $J_{ij}^{\alpha}$ is an exchange interaction between $S_i^{\alpha}$ and $S_j^{\alpha}$, $B$ is a biquadratic interaction, $\vec{h}$ is a magnetic field vector, and $D$ is the on-site spin anisotropy.
The model parameters and the spin quantum numbers are specified in the \verb|model| section of the input file (the default value of the spin quantum number is 0.5).
For a boson model, the Hamiltonian predefined in \tenessimple is given as
\begin{align*}
\mathcal{H} &= \sum_{\langle ij \rangle}\left[ -t_{ij} \left(b^\dagger_i b_j + b^\dagger_j b_i \right) + V_{ij} n_i n_j \right]\\
&+ \sum_i \left[U\frac{n_i(n_i-1)}{2} - \mu n_i\right],
\end{align*}
where $b_i^\dagger$ and $b_i$ are creation and annihilation operators of a boson at the $i$-th site, $n_i\equiv b_i^{\dagger}b_i$ is a number operator at the $i$-th site, $t_{ij}$ and $V_{ij}$ are a transfer integral and the Coulomb interaction between the $i$-th and $j$-th sites, respectively, $U$ is the onsite repulsion, and $\mu$ is the chemical potential. 
The model parameters and the maximum occupation number per site, $n_\text{max}$, are specified in the input file (the default value of the maximum occupation number is $1$).
In \tenessimple, two-body interactions up to third-nearest-neighbor sites can be treated. 

In the \verb|[lattice]| section, the type of lattice, the unit cell size in the $x$ and $y$ directions, the bond dimension, the initial state, and the noise for elements in the initial tensors are specified.
The type of lattice can be selected from square, triangle, honeycomb, and kagome lattices.
As an initial state, ferromagnetic, antiferromagnetic, and random states can be selected.
For a bosonic system, the ferromagnetic state indicates that $n_\text{max}$ particles are at all sites, while the antiferromagnetic state indicates that $n_\text{max}$ particles (zero particle) are on one sublattice (the other sublattice).

An example of an input file for the $S=1/2$ Heisenberg model on the square lattice is as follows:
\begin{verbatim}
[model]
type = "spin"       # type of model
J = 1.0             # Heisenberg int.
[lattice]
type = "square lattice"  # type of lattice
L = 2                    # size of unit cell
W = 2                    # size of unit cell
virtual_dim = 3          # bond dimension
initial = "antiferro"    # initial state
\end{verbatim}
For an input file \verb|simple.toml|, we can execute \tenessimple by the following command:
\begin{verbatim}
$ tenes_simple simple.toml    
\end{verbatim}
If \tenessimple is successfully executed, the output file \verb|std.toml| is generated.
A user can use this file as an input file for \tenesstd, which is described in the next section.

\subsection{Usage of \tenesstd}

The script \tenesstd is a tool for generating an input file for \tenes by calculating imaginary-time evolution operators $\exp(-\tau\ham_{ij})$ from a given Hamiltonian $\ham$ and an imaginary-time step $\tau$.
For a long range interaction, the imaginary-time evolution operator is decomposed into a product of nearest-neighbor operators according to the procedure described in Sec.~\ref{sec:alg_optim}.
Thus, for a lattice and/or a model which is not supported in \tenessimple, an input file of \tenesstd can be prepared by modifying a file generated by \tenessimple or prepared from scratch.
The input file of \tenesstd requires three sections, \verb|parameter|, \verb|tensor|, and \verb|hamiltonian|.

In the \verb|parameter| section, various parameters of the calculation such as the number of updates are specified.
This section has five subsections: \verb|general|, \verb|simple_update|, \verb|full_update|, \verb|ctm|, and \verb|random|.
In the \verb|general| subsection, we can specify general parameters for \tenes such as a name of the directory for saving (loading) tensors, a cutoff value for operator elements, and whether elements of tensors are restricted to real numbers or not.
In the \verb|simple_update| subsection, we can specify parameters in the simple update procedure such as an imaginary time step $\tau$ of the imaginary time evolution and the total number of simple updates. 
In the \verb|full_update| subsection, one can specify parameters in the full update procedure such as the imaginary time step $\tau$ and the cutoff of singular values.
In the \verb|ctm| subsection, we can specify parameters for corner transfer matrices (CTMs) such as the bond dimension $\chi$, a convergence criteria, and whether the randomized SVD method is used or not.
Additionally, whether the CTMRG method or the MFE method is used for contraction in the evaluating process can be specified in this subsection.
In the \verb|random| subsection, we can specify the seed of the pseudo-random number generator used in initializing the tensor.
The \texttt{parameter} section is copied into the input file for \tenes that is generated by \tenesstd.

In the \texttt{tensor} section, information about iTPS tensors on a square lattice\footnote{It should be noted again that TeNeS always considers iTPS on a square lattice.} are specified.
\texttt{L\_sub} specifies the dimensions of the unit cell.
In order to deal with skew boundary condition, \texttt{skew} can be used.
This value means the number of sites the unit cell is shifted along the x-axis by when moving one unit cell along the y-axis.
For example, the tensor network depicted by Figure~\ref{fig:iTPS} is prepared by \texttt{L\_sub = [2,1]} and \texttt{skew = 1}.
The \texttt{[tensor]} section has \texttt{[[unitcell]]} subsections.
These subsections specify the parameters of each site tensor in the unit cell.
For example, \texttt{physical\_dim} and \texttt{virtual\_dim} give the dimensions of the bonds.

In \tenesstd (as well as in \tenes), the Hamiltonian is composed of the two-body terms as $\ham = \sum_{\{(i, j)\}} \ham_{ij}$.
In the \verb|[[hamiltonian]]| sections of the \tenesstd input, we can specify $\ham_{ij}$.
Each \verb|[[hamiltonian]]| has three fields; \verb|dim| specifies the local degree of freedom at sites $(i,j)$, \verb|bonds| specifies the list of bonds $(i,j)$, and \verb|elements| specifies the matrix element of $\ham_{ij}$.
While this section can be generated easily by executing \tenessimple for simple models, we can design a quantum lattice model flexibly by defining the list of bonds and the matrix elements of the operators used in $H_{ij}$.
Note that we can specify an arbitrary lattice by modifying sites in a unit cell into a square or rectangular configuration and by connecting sites with bonds to reproduce the topology of the lattice under consideration.

As an example, let us define the Hamiltonian for the $S=1/2$ Heisenberg model on a square lattice with a $2\times 2$ unit cell (specified by \verb|L_sub = [2,2]| in the \verb|tensor| section).
The two-site term of the Hamiltonian is given as follows:
\begin{align}
\ham_{ij} = S_i^z S_j^z + \frac{1}{2} \left[S_i^+ S_j^- + S_i^- S_j^+ \right]\label{eq:heisenberg_std}.
\end{align}
In this case, the dimension of the operators $S_i^z$ and $S_i^\pm$ is set as \verb|dim = [2,2]| because the state at each site is represented by a basis set composed of the two states \(|\uparrow\rangle\) and \(|\downarrow\rangle\).
Each line of \verb|bonds| specifies, by three integers, a pair of sites on which the two-site operator acts, where the first integer is the index of a source site and the other two integers are the relative coordinates $(dx, dy)$ of the target site from the source site. Note that TeNeS can treat relative coordinates in the range $|dx|, |dy| \le 3$.
For example, in the case of the Heisenberg model with nearest-neighbor interactions on a square lattice, \verb|bonds| is specified as follows: 
\begin{verbatim}
bonds = """
0 0 1
0 1 0
1 0 1
1 1 0
2 0 1
2 1 0
3 0 1
3 1 0
"""
\end{verbatim}
Here, we have set the index number of the sites in a unit cell as shown in Fig.~\ref{fig:heisenberg_model}.
In the appendix \ref{appendix:kagome}, we show another instructive example -- how to represent a kagome lattice.

\begin{figure}
  \centering
  \includegraphics[width=0.5\columnwidth]{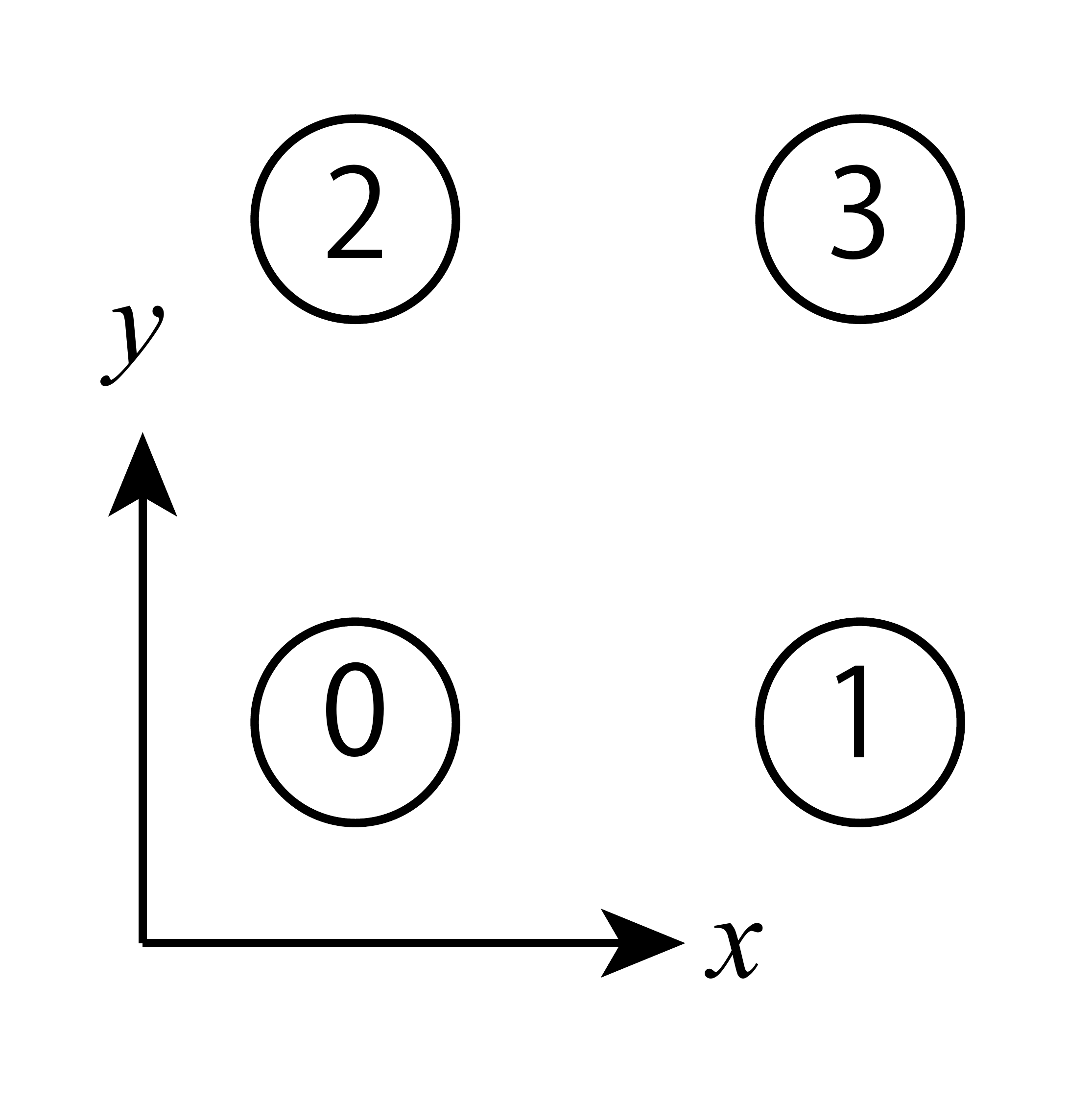}
  \cprotect\caption{Site indices of the $S=1/2$ Heisenberg model on a square lattice at \verb|Lsub=[2,2]|. }
  \label{fig:heisenberg_model}
\end{figure}

The matrix elements of the two-site operators are specified by \verb|elements|.
One matrix element of the operator can be specified by one line, composed of four integers and two floating-point numbers.
For example, the operator given in Eq.~(\ref{eq:heisenberg_std}) is specified as follows:
\begin{verbatim}
elements = """
0 0 0 0  0.25 0.0
1 0 1 0  -0.25 0.0
0 1 1 0  0.5 0.0
1 0 0 1  0.5 0.0
0 1 0 1  -0.25 0.0
1 1 1 1  0.25 0.0
"""    
\end{verbatim}
In each line, the first two integers represent the state of the source and target sites before the operator acts on them, the subsequent two integers represent those after the action, and the last two floating-point numbers indicate the real and imaginary parts of the matrix element of the operator.
For example, the first line indicates $\braket{ \uparrow_i \uparrow_j | \mathcal{H}_{ij} | \uparrow_i \uparrow_j } = 1/4$ where the states  $\ket{\uparrow}$ and $\ket{\downarrow}$ are labeled as $0$ and $1$.
Similarly, the second line indicates $\braket{\downarrow_i \uparrow_j | \mathcal{H}_{ij} | \downarrow_i \uparrow_j} = -1/4$.
For detailed information on customizing a lattice and a model, see the manual~\cite{TeNeS_webpage, TeNeS_Github}.

If we prepare the input file \verb|std.toml|, \tenesstd can be executed by the following command:
\begin{verbatim}
$ tenes_std std.toml    
\end{verbatim}
If \tenesstd is successfully executed, the output file \verb|input.toml| is generated.
This file can be used as an input file for \tenes, as described in the next section.

\subsection{Usage of \tenes}

\tenes is the main program of TeNeS.
\tenes needs an input file that defines the model and the operators appearing in it and specifies the updates, model parameters, and other factors.
The user can specify the observables and/or the numerical condition of the two-site correlation functions by directly editing the input to \tenes.
The input file of \tenes has the following sections: \verb|parameter|, \verb|tensor|, \verb|evolution|, \verb|observable|, \verb|correlation|, and \verb|correlation_length|. 

The \verb|parameter| section is almost the same as that in the input file of \tenesstd except that imaginary time steps $\tau$ for both simple and full updates are used no longer. The \verb|tensor| section is the same as in \tenesstd.
In the \verb|evolution| section, the imaginary time evolution operators used in simple and full updates are specified.

In the \verb|observable| section, the observables to be measured are specified. 
This section has two subsections, \verb|onesite| and \verb|twosite|, where one-body and two-body operators are respectively defined.
In the \verb|correlation| section, information for the two-site correlation functions to be measured can be specified.
The form of the correlation function is given as
\begin{align}
    C({\bm r}, {\bm r}_0) = \langle A({\bm r}_0) B({\bm r}_0+{\bm r})\rangle ,
    \label{eq:corr}
\end{align}
where ${\bm r}_0$ and ${\bm r}_0+{\bm r}$ are cartesian coordinates on a square lattice.
For example, the displacements ${\bm r} = (1,0)$ and ${\bm r} = (0,1)$ mean the right-neighbor and top-neighbor sites from ${\bm r}_0$, respectively.
\tenes calculates the correlation functions along the directions of the $x$ and $y$ axis, i.e., for the displacements
${\bm r} = (i, 0)$ and $(0,i)$ for $0\le i \le r_{\rm max}$, where $r_{\rm max}$ is a parameter specified in the \verb|correlation| section.
For a fixed ${\bm r}$, the two-body correlation function is calculated by
taking all the sites in a unit cell as the start-point site ${\bm r}_0$.
The operators $A$ and $B$ in Eq.~(\ref{eq:corr}) are specified in the \verb|observable.onesite| subsection.

In the \verb|correlation_length| section, the input parameters related to measuring the effective correlation length of bulk tensors are specified.
The effective correlation length $\xi$ is calculated as $\xi^{-1} = e_1 = -\log(\lambda_1/\lambda_0)$, where $\lambda_i$ is the $i$-th eigenvalue of the transfer matrix ($|\lambda_0| \ge |\lambda_1| \ge |\lambda_2| \ge \cdots$).
In addition, \tenes calculates more eigenvalues $e_2, e_3, \dots$ to help estimate a finite $\chi$ error of $\xi$ as~\cite{Rams2018, Rader2018}
\begin{align}
    \frac{1}{\xi(\infty)} &= \frac{1}{\xi(\chi)} + k \log\left|\frac{\lambda_1(\chi)}{\lambda_2(\chi)}\right| \nonumber \\
    & = \frac{1}{\xi(\chi)} + k(e_2(\chi) - e_1(\chi)).
\end{align}

If we prepare an input file \verb|input.toml|, \tenes can be executed by the following command:

\begin{verbatim}
$ tenes input.toml    
\end{verbatim}
If the execution successfully finishes, the following files are output in the output directory: 
\begin{itemize}
\item \verb|parameters.dat|: parameters in the \verb|parameter| and \verb|lattice| sections given in the input file, 
\item \verb|density.dat|: expected values per site of each observable,
\item \verb|onesite_obs.dat|: expected values of the site operator,
\item \verb|twosite_obs.dat|: nearest-neighbor-site correlations for site operations, 
\item  \verb|correlation.dat|: correlation functions, 
\item  \verb|correlation_length.dat|: correlation lengths and eigenvalues of transfer matrices, 
\item  \verb|time.dat|: calculation time.
\end{itemize}

If \texttt{tensor\_save} parameter in the \texttt{[parameter.general]} section is defined,
the obtained tensors will be saved into a directory whose name is specified by the \texttt{tensor\_save}.
These tensors can be loaded as an initial state in another calculation by setting \texttt{tensor\_load} parameter in the \texttt{[parameter.general]} section.
By using this mechanism, users can re-calculate additional observable for the given
states without the optimization process.
Another usage of this mechanism is to prepare a proper initial state for a system whose parameters differ a little.

See the TeNeS manual\cite{TeNeS_webpage, TeNeS_Github} for more information. 

\section{Application Examples}
\label{sec:applications}
In this section, we present a few examples that illustrate the use of TeNeS. 
In each example, the toml input file for \tenessimple is provided in the TeNeS package together with a script for running \tenessimple, \tenesstd, and \tenes, so a user can also generate the input files for \tenesstd and \tenes as well, simply by running the script, which provides a good starting point for further customization to meet various needs.

\subsection{Transverse-field Ising model}
\label{sec:TransverseIsing}

\begin{figure}[tb!]
  \begin{center}
    \includegraphics[width=.9\columnwidth]{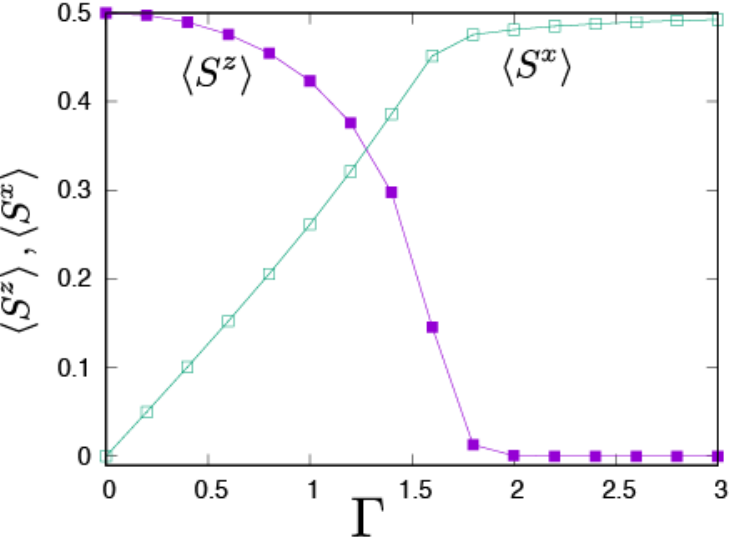}
    \cprotect\caption{Calculated values of $\langle S^z\rangle$ and $\langle S^x \rangle$ for the two-dimensional transverse Ising model shown as a function of the transverse magnetic field $h_x$. In this calculation, we set the number of simple and full updates as $1000$ and 0, the imaginary time as $\tau = 0.01$, the maximum iteration number of convergence for CTM as $10$, the bond dimension as $D=2$, and the bond dimension of CTM as $\chi=10$.
    }
    \label{fig:TransverseIsing}
  \end{center}
\end{figure}

Let us first consider the two-dimensional transverse-field Ising model as a simple example.
The Hamiltonian of this model is given as
\begin{align}
\ham = J_z \sum_{\langle i,j \rangle} S_i^z S_j^z - h_x \sum_i S_i^x.
\end{align}
In this example, we consider a ferromagnetic exchange interaction ($J_z = -1$).
A sample input file, \verb|simple.toml|, is prepared in the \verb|sample/01_transverse_field_ising| directory.
TeNeS can be executed in the simple mode by the following command:
\begin{verbatim}
$ tenes_simple simple.toml
$ tenes_std std.toml
$ tenes input.toml
\end{verbatim}
If the execution is successful, several output messages are displayed.
They include information on the parallelization, which tensors used in calculation are real or complex, execution status of the calculation process, and execution wall times.
This output message shows the expectation value of the one-site spin operators $\langle S_i^z \rangle$ and $\langle S_i^x \rangle$, the energy $\langle \ham\rangle$, and two-site correlation functions $\langle S^x_i S^x_j \rangle$, $\langle S^y_i S^y_j \rangle$, and $\langle S^z_i S^z_j \rangle$ for neighboring sites $\langle i,j \rangle$ (the two numbers for each observable are the real and imaginary parts, respectively).
The execution of TeNeS can be repeated by changing the magnetic field $h_x$ using the Python3 codes prepared in the same directory.
They are executed from the terminal as follows:
\begin{verbatim}
$ python tutorial_example.py
$ python tutorial_read.py > output.dat
\end{verbatim}
The first command that executes calculation will finish within one minute.
Then, an array of $(h_x,\langle\ham\rangle,\langle S^z_i \rangle, \langle S^x_i \rangle)$ is output into the \verb|output.dat| file, which can be plotted using software such as \verb|gnuplot|.
Fig.~\ref{fig:TransverseIsing} shows the calculated $\langle S^z_i \rangle$ and $\langle S^x_i \rangle$ as functions of $h_x$. 

\subsection{Heisenberg model}
\label{sec:Heisenberg}

In this subsection, the calculation of the magnetization curves of the $S=1/2$ quantum Heisenberg model on square and triangular lattices is demonstrated as an example. 
The Hamiltonian is given as
\begin{align}
\ham = J \sum_{\langle i,j \rangle}\sum_{\alpha}^{x,y,z} S_i^\alpha S_j^\alpha - h \sum_i S_i^z,
\end{align}
where $h$ represents an external magnetic field and $J$ ($>0$) is an antiferromagnetic exchange coupling.
Using \tenes, we calculate
the magnetization per site, $\langle S^z \rangle \equiv \sum_i^{N_\mu} \langle S_i^z \rangle/N_\mu$, as a function of $h$, where $N_{\mu}$ is the total number of sites in the unit cell.
In the \verb|sample/05_magnetization| directory, samples of the input files, \verb|basic.toml| and \verb|basic_square.toml|, are prepared, corresponding to the triangular and square lattice, respectively.
With these input files, TeNeS can calculate the magnetization using the simple mode.
These input files use only a simple update to reduce the computational time.
We note that a simulation using a full update will give more accurate results at the expense of computational time.
In the same directory, a Python3 code, \verb|tutorial_magnetization.py|, is prepared for obtaining a magnetization curve by repeated execution of TeNeS changing the magnetic field $h$.

\begin{figure}[tb!]  \begin{center}
    \includegraphics[width= 0.9\columnwidth]{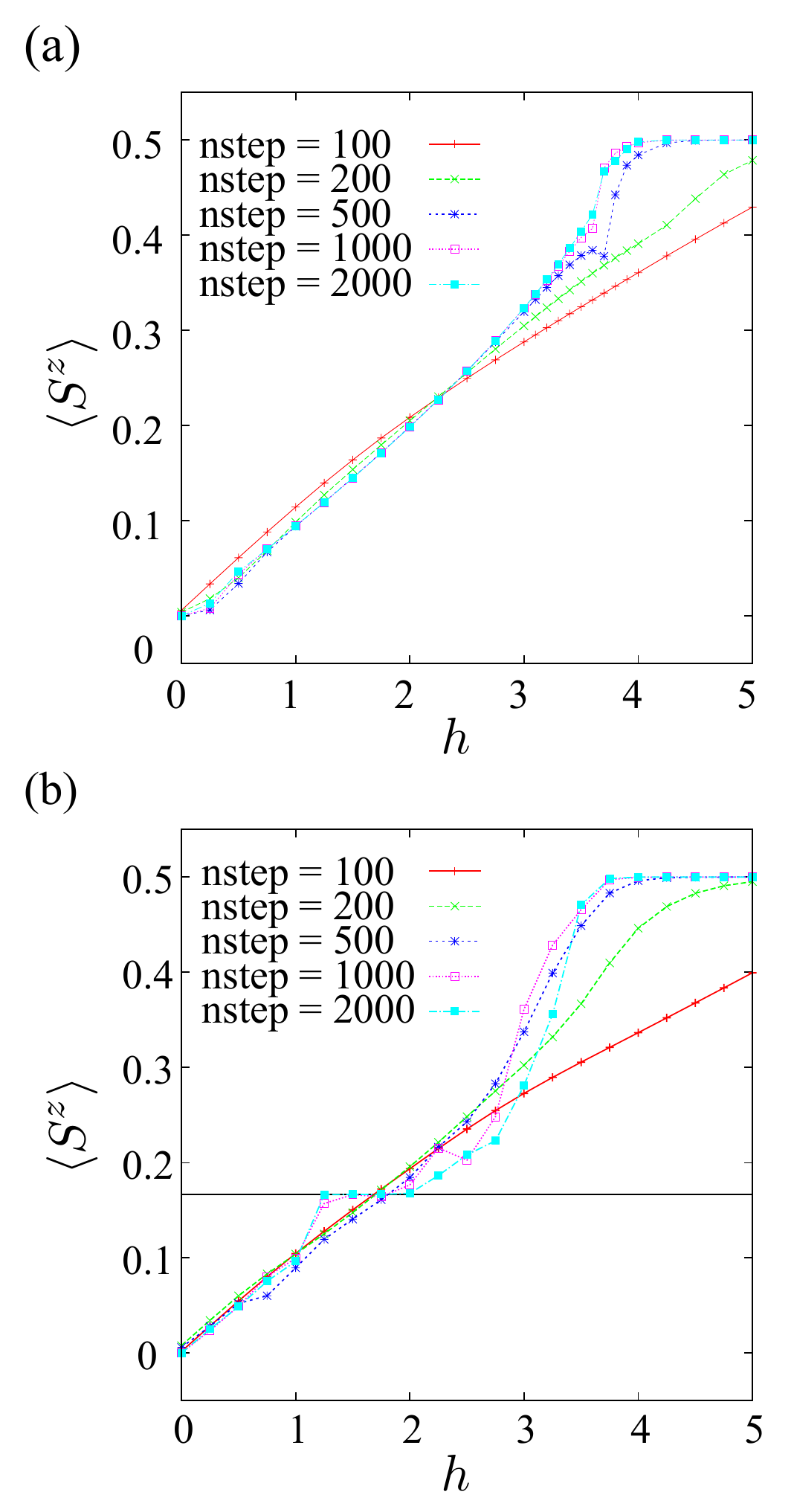}
    \cprotect\caption{Magnetic field $h$ dependence of the $\langle S^z \rangle$ of the Heisenberg model on (a) the square lattice and (b) the triangular lattice.
    The number ``nstep'' indicates the number of the simple update.
    The horizontal line corresponds to the ``1/3-plateau'' characteristic to the geometric frustration of the triangular lattice.
    In this calculation, we set the number of full updates as $0$, the imaginary time as $\tau = 0.01$, the maximum iteration number of convergence for CTM as $100$, the bond dimension as $D=2$, and the bond dimension of CTM as $\chi=10$.
    }
    \label{fig:tutorial_5}
  \end{center}
\end{figure}

Figures \ref{fig:tutorial_5} (a) and (b) show $h$ dependence of $\langle S^z \rangle$ for the Heisenberg model on square and triangular lattices, respectively. 
With increasing number of simple updates, the magnetization curve seems to converge.
In the case of the triangular lattice, it is known that a characteristic magnetic structure with $\langle S^z \rangle = 1/6$ appears in the intermediate magnetic field region \cite{QuantumMagnetism,Honecker_2004}.
To see it clearly, a straight line $\langle S^z \rangle = 1/6$ is drawn in Fig. \ref{fig:tutorial_5} (b).
We can see that a $\langle S^z \rangle = 1/6$ plateau structure appears in the magnetic-field region $1.2 \lesssim h \lesssim 2.0$ when the number of simple updates (nstep) is set as $2000$. 

\subsection{Hardcore boson model}
\label{sec:HardcoreBoson}

\begin{figure}
  \centering
  \includegraphics[width=\columnwidth]{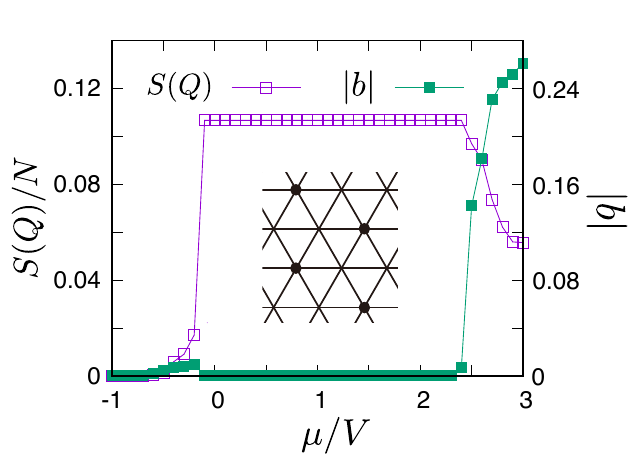}
  \caption{Two kinds of order parameters of the hardcore boson model on a triangular lattice plotted as a function of the chemical potential.
  The hopping is taken as $t/V=0.1$.
  The inset shows an ordering pattern in a solid phase.
  In this calculation, we set the number of simple updates and full updates as $3000$ and $0$, the imaginary time as $\tau = 0.01$, the maximum iteration number of convergence for CTM as $3000$, the bond dimension as $D=5$, and the bond dimension of CTM as $\chi=25$.
  }
  \label{fig:hardcoreboson}
\end{figure}

Finally, let us consider a hardcore boson model on a triangular lattice as an example of a bosonic system.
The Hamiltonian of this model is given as
\begin{align}
\ham &= \sum_{\langle i,j \rangle} \Bigl[ -t (b_i^\dagger b_j + b_j^\dagger b_i) + V n_i n_j \Bigr] -\mu \sum_i n_i, 
\end{align}
where $t$ is the hopping energy, $V$ is the Coulomb interaction between neighboring sites, and $\mu$ is the chemical potential.
We assume a hardcore boson system, i.e., a system in which the maximum number of bosons at each site is 1.
It should be noted that this system is invariant under the particle--hole transformation, $\mu \leftrightarrow 6V-\mu$,
and hence it is sufficient to calculate up to $\mu/V = 3$.
The Python3 code for simulation of the hardcore boson model is prepared in the \verb|sample/06_hardcore_boson_triangular| directory.
In this model, several phases characterized by two types of long-range order are expected to be realized~\cite{Wessel2005}.
One is a solid-like order which exists at a $1/3$ and $2/3$ filling, where one of three sites is filled in a $\sqrt{3}\times\sqrt{3}$ ordering with wave vector ${\bm Q}=(4\pi/3,0)$ (see the inset of Fig.~\ref{fig:hardcoreboson}).
This long-range order is characterized by the structure factor $S({\bm Q})$.
The other is a superfluid order which is characterized by the off-diagonal order parameter $\langle b \rangle$.
Fig.~\ref{fig:hardcoreboson} shows the two order parameters as a function of the chemical potential.
The figure indicates that there exist three phases: (a) a superfluid phase ($-0.5 \lesssim \mu/V \lesssim -0.2$), (b) a solid phase ($-0.2 \lesssim \mu/V \lesssim 2.4$), and (c) a supersolid phase, where both orders appear simultaneously ($2.4 \lesssim \mu/V$).
This result is consistent with a previous numerical simulation~\cite{Wessel2005}.

\section{Summary}
\label{sec:summary}

In this paper, we introduced a FLOSS package, TeNeS, which searches for ground states of two-dimensional quantum lattice models with tensor network techniques based on iTPS.
We described the techniques implemented in TeNeS and how to install and use it.
We also presented a few applications.

While TeNeS may not be the most efficient tensor-network-based solver of every specific problem at hand,
it aims to be a general solver for various lattice models by employing as simple an algorithm and a scheme as possible.
For example, a decomposition of a long-ranged ITE tensor into successive nearest-neighbor ITE tensors on a square lattice gives TeNeS the flexibility to deal with an arbitrary lattice in a simple way.
We are planning to implement more algorithms/methods to TeNeS;
calculation of fermionic systems,
finite temperature calculation,
treatment of symmeties,
gradient-based optimization using auto-differentiation,
and so on.

Since TeNeS is a ready-to-use solver for quantum lattice systems, numerical calculation using tensor networks can be performed easily even by researchers who are not experts.
TeNeS will also help experts of computational condensed matter physics for testing and developing new algorithms and programs.
We hope that TeNeS will lower the technical barrier for tensor network calculations and will stimulate the research activity of this developing field.

\section*{Acknowledgments}
TeNeS was developed under the support of the ``Project for advancement of software usability in materials science'' (PASUMS) in fiscal year 2019 by The Institute for Solid State Physics, The University of Tokyo. This work was partially supported by MEXT as ``Exploratory Challenge on Post-K computer'' (Frontiers of Basic Science: Challenging the Limits), by JSPS KAKENHI Nos.~15K17701 and 19K03740, and by JST PRESTO Grant No. JPMJPR1912. TO acknowledges support by the Endowed Project for Quantum Software Research and Education, The University of Tokyo (https://qsw.phys.s.u-tokyo.ac.jp/).
The computations in this work were done using the facilities of the Supercomputer Center, The Institute for Solid State Physics, The University of Tokyo.

\appendix

\section{Example of non-trivial lattice -- kagome lattice}\label{appendix:kagome}

In this appendix, we briefly described how to implement another lattice by using an example, kagome lattice
\footnote{Note that the simple mode \tenessimple can generate an input file of the standard mode for the kagome lattice.}.
First, we should represent the lattice by using a square lattice.
For example, the kagome lattice can be embedded into the square lattice by adding a ``dummy'' site (Fig.~\ref{fig:kagome}).
To reduce the computational cost and the memory cost, the dimensions of the physical bond and the four virtual bonds of the tensor on the dummy site can be set to one.
Thus, the unitcell is defined as follows:
\begin{verbatim}
[[tensor.unitcell]]
index = [0]
virtual_dim = [2,2,2,2]
physical_dim = 2

[[tensor.unitcell]]
index = [1]
virtual_dim = [2,1,2,1]
physical_dim = 2

[[tensor.unitcell]]
index = [2]
virtual_dim = [1,2,1,2]
physical_dim = 2

[[tensor.unitcell]]
index = [3]
virtual_dim = [1,1,1,1]
physical_dim = 1
\end{verbatim}
where the dimensions of the physical bond and the virtual bonds are 2.

Next, we define the bond Hamiltonians.
From Fig.~\ref{fig:kagome}, the unit cell has 6 bonds;
(1) the site 0 and the right site 1,
(2) the site 0 and the top site 2,
(3) the site 1 and the top-left site 2,
(4) the site 1 and the right site 0,
(5) the site 2 and the top site 0, and
(6) the site 2 and the top-left site 1.
These are specified as follows:
\begin{verbatim}
bonds = """
0 1 0
0 0 1
1 -1 1
1 1 0
2 0 1
2 -1 1
"""
\end{verbatim}

\begin{figure}
    \centering
    \includegraphics[width=.7\linewidth]{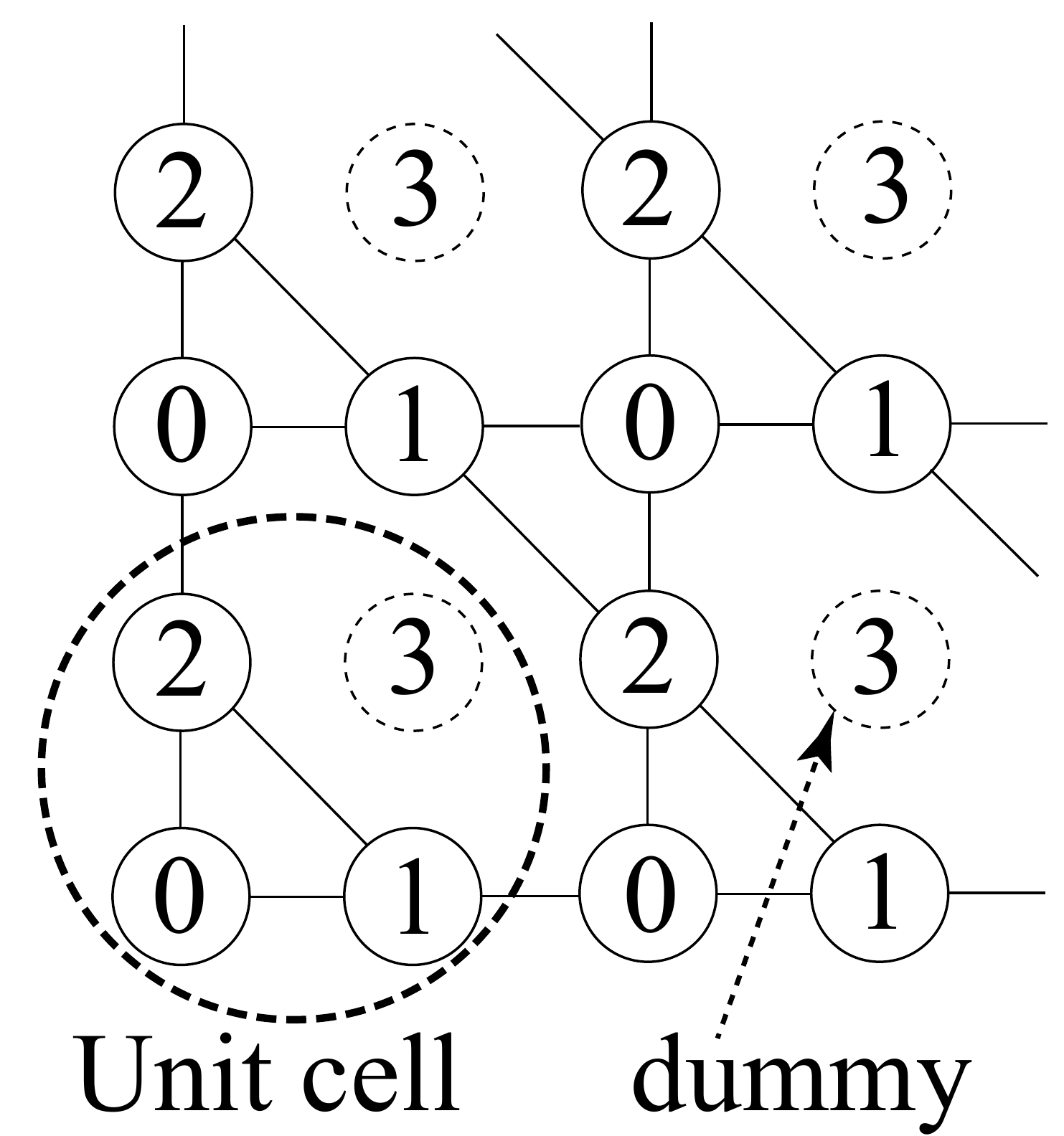}
    \caption{A kagome lattice represented on a square lattice. Sites indexed as ``3'' are dummy sites.}
    \label{fig:kagome}
\end{figure}

\bibliographystyle{elsarticle-num}
\bibliography{main}

\end{document}